\RequirePackage{fix-cm}
\documentclass[smallextended]{svjour3}       
\smartqed  
\usepackage[utf8]{inputenc}
\usepackage{graphicx}
\usepackage{amsmath}
\usepackage{ulem}
\usepackage[numbers,sort&compress]{natbib}

\usepackage{color}

\begin{document}

\title{Residual cut-off dependence and power counting: \\ the deuteron as a case study
}
\subtitle{Knowns, unknowns and known unknowns about regulator dependence}


\author{Daniel~Odell         \and
  Manuel~Pavon~Valderrama \and
  Lucas~Platter}


\institute{D. Odell \at
           Department of Physics and Astronomy, Ohio University, Athens, OH 45701, USA;
           Department of Physics and Astronomy, University of Tennessee, Knoxville, TN 37996, USA;
  \email{dodell@ohio.edu} 
           \\
           M. Pavon Valderrama \at
           School of Physics, Beihang University, Beijing 100191, China;            \email{mpavon@buaa.edu.cn}
           \\
           L. Platter \at
           University of Tennessee, Knoxville, Department of Physics and Astronomy,
           Knoxville, TN 37996, USA;  Oak Ridge National Laboratory, Physics Division, Oak Ridge, TN, USA;
           \email{lplatter@utk.edu}
}

\date{Received: date / Accepted: date}

\maketitle

\begin{abstract}
  Effective field theories (EFTs) require regularization and renormalization
  to gain predictive power.
  While regularization is inconsequential from the point of view of
  the observable predictions of EFT --- in a renormalized theory
  we expect predictions to be regulator-independent
  once the cutoff is removed --- the particular details of regulator dependence
  might provide interesting insights into the inner workings of an EFT.
  In fact, the analysis of regulator dependence has been frequently suggested
  as a tool to study the ordering scheme or {\it power counting} of EFTs.
  We show here that the choice of the regulator might impact the power law
  properties of the residual cutoff dependence.
  If this conclusion were to be confirmed, it would have consequences
  on the validity of this method as a tool to analyze power counting.
\end{abstract}

\section{Introduction}
\label{intro}
The nuclear interaction, despite having been studied for decades,
is still not completely understood at the theoretical level.
After Yukawa's seminal proposal of the pion, the central role of meson exchanges
was eventually recognized and models including the exchange of different
types of mesons were constructed.
These meson-exchange potentials were indeed able to describe
nucleon-nucleon scattering to a very high accuracy~\cite{Machleidt:2017vls}.

However, the eventual discovery of quantum chromodynamics (QCD) relegated
these meson theories to mere phenomenological descriptions of
the nuclear force.
Naturally a QCD-based derivation of the nuclear force was required
at this point, yet QCD is not analytically solvable at low energies,
leading to a theoretical deadlock.
In this regard the effective field theory (EFT) approach allows for an indirect
solution of QCD, where Weinberg suggested the first EFT describing
the nucleon-nucleon interaction~\cite{Weinberg:1990rz,Weinberg:1991um}.
This EFT incorporates long-range pion exchange contributions, which are
constrained by chiral symmetry, and contact-range interactions,
which systematically account for the unknown short-range
contributions.
In this approach the pion, the lightest meson, plays a special role:
the reason why the pion is lighter is chiral symmetry, which is
the main low-energy manifestation of QCD.
The effects of all other mesons are then lumped together in the contact
interactions and are thus treated as effects that cannot be resolved
at low-energies.

Yet, for EFTs to have predictive power, a {\it power counting} is required,
{\it i.e.} a principle that orders interactions from most to least relevant.
It happens that when Weinberg first proposed his scheme, the power counting
of non-perturbative EFTs was not well understood. But nuclear physics is
non-perturbative, which poses a problem regarding how to formulate
a power counting.
Weinberg's workaround was inspired: we can simply apply the standard,
perturbative power counting rules to the N-body potential, which is
after all the perturbative sum of irreducible diagrams.
Then we solve this potential non-perturbatively with the Schr\"odinger or
Lippmann-Schwinger equation, as has always been done
in nuclear physics.

This approach --- known as {\it Weinberg counting} --- has been very successful
phenomenologically, resulting in very good descriptions of
the nuclear force~\cite{Epelbaum:2014sza,Entem:2017gor}.
But the theoretical soundness of the Weinberg counting has been put into
question, too~\cite{Kaplan:1996xu,Nogga:2005hy}, leading to a heated
debate on whether it is a correct EFT or not~\cite{Epelbaum:2018zli,PavonValderrama:2019uzi,Epelbaum:2019msl,PavonValderrama:2019lsu,Machleidt:2020vzm,vanKolck:2020llt}.
If anything this situation makes the analysis of power counting a particularly
relevant theoretical tool for this debate.
In this regard residual cutoff dependence has been argued to determine
the power counting of the subleading order corrections and thus it
deserves
attention~\cite{Long:2011qx,Long:2011xw,Long:2012ve,Griesshammer:2015osb,Valderrama:2016koj,Griesshammer:2020fwr}.
The nuclear interaction derived from chiral EFT needs
to be regularized with a short-distance (or large momentum) cutoff and
residual cutoff dependence is expected to decrease as more orders in
the EFT expansion are included.
Besides, residual cutoff dependence might furthermore be important
for gaining insight into the EFT truncation related uncertainties
of observables.

However for this tool to be useful it is necessary that the power-law
properties of the residual cutoff dependence are regulator-independent.
If this were not the case, the conclusions derived from residual cutoff
dependence would be inconclusive.
In this contribution we find that there might be reasons to think that
the details of residual cutoff dependence might be dependent
on the regulator.
Owing to numerical complexities, it is difficult to give a definite answer
to this question, which will undoubtedly require a more thorough analysis
than the one we present here.
Nonetheless, it is interesting to note that residual cutoff dependence
is not obviously equivalent for different regulators.

We will focus here on the leading order of the chiral EFT interaction
-- the one-pion exchange potential -- and consider it in the
spin-triplet channel. We will regularize and renormalize it using five different
techniques and analyze the residual cutoff dependence.

This work is organized as follows: in
Sec.~\ref{sec:pionless-eft:-poster}, we discuss the {\it naive
  expectation} for regulator dependence in EFTs using the example of
the so-called pionless EFT. We introduce the one-pion exchange
potential in the following section, explain our approaches to
calculating observables and give and compare results for the different
regulators that we consider in this work. We conclude with a summary
in Sec.~\ref{sec:summary}.

\section{Pionless EFT: the poster child of residual cutoff dependence}
\label{sec:pionless-eft:-poster}
Pionless EFT is the effective field theory whose only degrees of
freedoms are nucleons~\cite{vanKolck:1998bw, Kaplan:1998we, Kaplan:1998tg}.
As there are no pions in this EFT, nucleons interact through contact
interactions only.
It is the quantum field theoretical formulation of
effective range theory and therefore produces the effective range
parameters such as the scattering length and effective range in the
two-body sector, while also accounting for many-body interactions and
many-body currents when needed to include necessary short-distance
physics.
It has been successfully applied to various few-body systems and
has been used to calculate various electroweak observables
to high accuracy.

We will use pionless EFT to illustrate the relation between residual
cutoff dependence and power counting.
This is particularly practical as the power counting of pionless EFT is very
well understood, in contrast with what happens in pionful EFT.
For simplicity, let us consider two-component fermions:
the pionless Lagrangian that describes the dynamics of
such a system is
\begin{eqnarray}
  \label{eq:Lagrangian-pionless}
\nonumber
  \mathcal{L}&=&
\sum_{\sigma=1,2}\psi_\sigma^\dagger\left(i\partial_t+\frac{\nabla^2}{2m}\right)\psi_\sigma
-C_{0}\psi_1^\dagger\psi_2^\dagger\psi_2\psi_1
+\frac{C_{2}}{8}\left(\psi_1^\dagger\overleftrightarrow{\nabla}^{2}\psi_2^\dagger\psi_2\psi_1+\rm{h.c.}\right)+\ldots,
\end{eqnarray}
where for simplicity we have only included the two lowest order operators:
the operators required to reproduce the scattering length and
the effective range correction in the two-body system.
If the scattering length is unnaturally large, the power counting that this
system follows is well known~\cite{vanKolck:1998bw,Chen:1999tn}:
the $C_0$ non-derivative interaction enters
at leading order (${\rm LO}$), while the $C_2$ derivative interaction is
suppressed by one order in the EFT expansion (i.e. by one power of $Q/M$)
and is next-to-leading order (${\rm NLO}$) then.
Later we will see how this is reproduced from analyzing the residual cutoff
dependence.

The coefficients in the Lagrangian will be determined by demanding that
the scattering amplitude calculated within the EFT reproduces
the on-shell amplitude obtained with
the effective range expansion
\begin{equation}
  \label{eq:tmatrix}
  t(k) = \frac{4\pi}{m_N} \frac{1}{-1/a + r_0/2 k^2 + \dots  - i k} \approx
  \frac{4\pi}{m_N}\frac{1}{-\frac{1}{a}-ik}\left(1-\frac{\frac{r_0}{2} k^2}{-\frac{1}{a}-ik}+\ldots\right) \, .
\end{equation}
Specifically, we will aim at reproducing the first term
in the expanded form of the effective range amplitude
shown above. This reflects that we will consider the zero-range limit
in which the scattering length is
assumed to be large compared to the range of the interaction. The
operator with the coefficient $C_0$ will thus be included at leading
order (LO) of the calculation, and the operator with coefficient $C_2$
at next-to-leading order (NLO).

\begin{center}
  \begin{figure*}
    \centerline{
    \includegraphics[width = 0.8\textwidth]{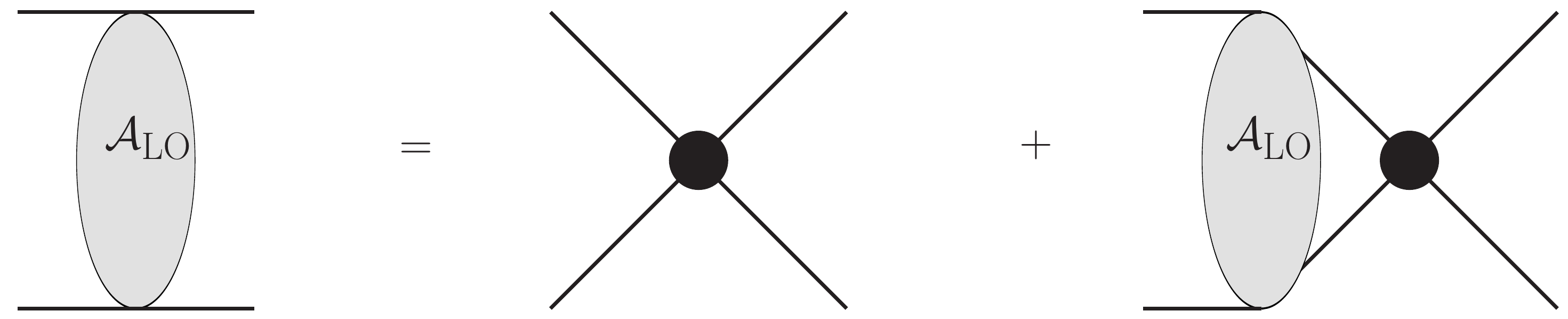}}
  \caption{Leading order scattering amplitude in the pionless EFT.}
  \label{fig:ALO}
\end{figure*}
\end{center}

The leading order EFT amplitude is evaluated by calculating the
diagrammatic integral equation in Fig.~\ref{fig:ALO} that leads to
\begin{equation}
  \label{eq:ALO_EFT}
  i\mathcal{A}_{\rm LO}(k)=-i
  C_{0}-iC_{0}\mathcal{I}_0(k,\Lambda)\, i\mathcal{A}_{\rm LO}(k)~,
\end{equation}
where $k$ denotes the relative momentum in the two-body system. The
function $\mathcal{I}_0(k,\Lambda)$ is the solution to the loop
integral in Fig.~\ref{fig:ALO} and depends on the momentum $k$ and the
ultraviolet cutoff $\Lambda$ that is imposed on the
integral.

The unregularized loop integral $\mathcal{I}_0$  is given by
\begin{eqnarray}
  \mathcal{I}_0(k) = i \int \frac{d^3 \vec{q}}{(2 \pi)^2}\,
  \frac{m_N}{k^2 + i \epsilon - q^2} \, ,
\end{eqnarray}
with $\vec{q}$ the loop momentum and $m_N$ the nucleon mass.
From direct inspection, it is apparent that this integral diverges
logarithmically for large loop momenta.
There are multiple ways to regularize the loop integral,
where here for concreteness we will consider cutoff
regularization, i.e.\ we will regularize the loop integral by including
a regulator function $\rho$ in the loop momentum, leading to
\begin{eqnarray}
  \mathcal{I}_0(k, \Lambda) = i \int \frac{d^3 \vec{q}}{(2 \pi)^2}\,
  \frac{m_N}{k^2 + i \epsilon - q^2} \,\rho^2(\frac{q}{\Lambda}) \, .
\end{eqnarray}
The integral is now finite, with its exact evaluation depending
on the regulator choice, where its general form will be
\begin{equation}
  \label{eq:I0loop}
  \mathcal{I}_0 \approx -\frac{i m }{ 2\pi^2}
  \,\left(\beta_0\,\Lambda+\frac{i \pi}{2}\sqrt{m_N E} -
  \beta_1\,\frac{m_N E}{\Lambda} + 
  \ldots\right) ~,
\end{equation}
where $\beta_0$ and $\beta_1$ are regulator-dependent numbers
(for instance, a sharp cutoff regulator will have $\beta_0 = \beta_1 = 1$)
and the ellipses denotes higher powers of $1/\Lambda$.
Now we can determine the coupling $C_0$ from the condition of reproducing
the scattering length, yielding
\begin{eqnarray}
  \frac{1}{C_0(\Lambda)} &=& \frac{m_N}{4 \pi}\,
  \left( \frac{1}{a_0} - \beta_0 \frac{2}{\pi} \Lambda\right) \, .
\end{eqnarray}
If the scattering length is large, it is counted as $a_0 \sim 1/Q$,
which leads to $C_0 \sim 1/Q$.
From this coupling we get
\begin{eqnarray}\
  k\,{\cot{\delta_{\rm LO}}} = -\frac{4\pi}{m_N C_0(\Lambda)} + 
  \frac{2}{\pi}\,\left( -\beta_0\,\Lambda +
  \beta_1\,\frac{k^2}{\Lambda} + \ldots \, ,
  \right)
\end{eqnarray}
or in terms of the $t$-matrix
\begin{eqnarray}
  t_{\rm LO}(k,\Lambda) &=&
  -\frac{4 \pi}{m_N}\,\frac{1}{k\,\cot{\delta_{\rm LO}} - i k} \, ,
\end{eqnarray}
from which the cutoff dependence turns out to be
\begin{eqnarray}
  t_{\rm LO}(k,\Lambda) &&=
  \frac{4 \pi}{m_N}\,\frac{1}{\frac{1}{a_0} + i k} +
  \frac{8 k^2}{\Lambda m_N}\,\frac{\beta_1}{{(\frac{1}{a_0} + i k)}^2} +
  \mathcal{O}(\frac{k^4}{\Lambda^2}) \, .
\end{eqnarray}
The residual cutoff dependence is proportional to $1/\Lambda$, i.e.
\begin{eqnarray}
  t_{\rm LO}(k,\Lambda) - \lim_{\Lambda \to \infty}\,t_{\rm LO}(k, \Lambda) \propto
  \mathcal{O}(\frac{1}{\Lambda}) \, ,
\end{eqnarray}
which indicates that the cutoff uncertainty is subleading by one order
in the EFT expansion, as expected.
As a consequence the contact-range operator removing this dependence
appears one order below the $C_0$ operator.
That is, $C_2$ should be ${\rm NLO}$, in agreement with the standard
power counting of pionless EFT.
Other types of regulators yield the same conclusions, including
power divergence subtraction (PDS)~\cite{Chen:1999tn}, local
regulators and boundary conditions~\cite{Valderrama:2016koj}, and hard
momentum space regulators~\cite{Emmons:2020aov}.
More elaborated demonstrations and examples of what can be done with residual
cutoff dependence can be found
in Refs.~\cite{Griesshammer:2015osb,Griesshammer:2020fwr}.

\section{One pion exchange}

Now we will turn our attention to the pionful theory, for which residual
cutoff dependence happens to be a considerably more involved issue.

The one-pion exchange potential is the leading order interaction
derived from chiral effective field theory. In momentum space, its form is
\begin{equation}
  \label{eq:OPEmomspace}
  V(\vec{q}) = -\frac{1}{(2\pi)^3} \left(\frac{g_A}{2f_{\pi}}\right)^2
  \vec{\tau}_1\cdot\vec{\tau}_2 \frac{ (\vec{\sigma}_1\cdot\vec{q})
  (\vec{\sigma}_2\cdot\vec{q}) }{ \vec{q}^2 + m_\pi^2  } \, ,
\end{equation}
and in coordinate space it is
\begin{equation}
  \label{eq:OPEcoordspace}
  V(\vec{r}) = \frac{m_\pi^2}{12\pi}
  \left(\frac{g_A}{2f_\pi}\right)^2(\vec{\tau}_1\cdot\vec{\tau}_2)
  \left[S_{12}~T(r) + (\vec{\sigma}_1\cdot\vec{\sigma}_2)Y(r)\right] \, .
\end{equation}
A key feature of the OPE potential is that it conserves spin $S$ and
total angular momentum $J$ but not angular momentum $L$.

\subsection{Description of the deuteron}
\label{sec:theory}

\paragraph{Coupled Channel Lippmann-Schwinger Equation:}
The momentum space Lippmann-Schwinger equation is written as a coupled
integral equation.
\begin{equation}
  t_{\alpha\alpha^\prime}(p,q_0) =
  \tilde{V}_{\alpha\alpha^\prime}(p,q_0) + \sum_{\alpha^{\prime\prime}}
  \int_0^\infty dq\,q^2\,\frac{\tilde{V}_{\alpha\alpha^{\prime\prime}}
  (p,q)t_{\alpha^{\prime\prime}\alpha^\prime}(q,q_0)}{E+i\epsilon-q^2/m}~.
\end{equation}
The OPE potential is not diagonal in angular momentum and couples
orbital angular momentum $l=0$ to $l=2$.
These coupled equations must be solved simultaneously.
To extract the spin-triplet scattering length from the zero-energy amplitude, we
calculate the effective range function via the Blatt-Biedenharn parameterization
according to \cite{deSwart:1995ui}.

\paragraph{Coupled Channel Schr\"odinger Equation:}
To describe the deuteron in r-space, we write the deuteron wave function as
\begin{eqnarray}
  \Psi_d(r; 1 m_d) = \frac{u(r)}{r}\,\mathcal{Y}^{01}_{1 m_d}(\hat{r}) +
  \frac{w(r)}{r}\,\mathcal{Y}^{21}_{1 m_d}(\hat{r}) \, ,
\end{eqnarray}
with $u$, $w$ the S- and D-wave reduced wave functions, $(j m) = (1 m_d)$
denoting the total angular momentum state of the deuteron and
$\mathcal{Y}^{ls}_{jm}$ the generalized spherical harmonics,
defined as
\begin{eqnarray}
  \mathcal{Y}^{ls}_{jm}(\hat{r}) = \sum_{m_l m_s} Y_{l m_l}(\hat{r}) | s m_s \rangle
  \langle l m_l s m_s | j m \rangle \, ,
\end{eqnarray}
where $\langle l m_l s m_s | j m \rangle$ is a Clebsch-Gordan coefficient.

If we write the potential in the deuteron channel as
the sum of a ``central'' and tensor pieces
\begin{eqnarray}
  V(r) = V_C(r) + V_T\,S_{12}(\hat{r}) \, ,
\end{eqnarray}
then the reduced wave functions obey the reduced Schr\"odinger equation
\begin{eqnarray}
  -u''(r) + m_N V_C(r) u(r) + m_N 2 \sqrt{2} V_T(r) w(r) &=& k^2\,u(r)
  \, , \nonumber \\ \\
  -w''(r) + m_N 2\sqrt{2}\,V_T(r) u(r) + \left[
    m_N (V_C - 2 V_T(r)) + \frac{6}{r^2} \right] w(r) &=& k^2\, w(r) \, .
  \nonumber \\
\end{eqnarray}

This equation can be solved by standard means.
If we are interested in the deuteron in particular we can simply set
$k^2 = -\gamma^2 = - m_N B_d$, with $B_d \approx 2.2$
the deuteron binding energy.

If we are interested in the scattering lengths, we set $k^2 = 0$ instead.
In this case there exists two linearly independent asymptotic ($r \to \infty$)
solutions, which we can call $\alpha$ and $\beta$:
\begin{eqnarray}
  u^{(\alpha)}(r) &\to&  1 - \frac{r}{a_0}~ ,\quad w^{(\alpha)}(r) \to \frac{a_{02}}{a_0} \frac{3}{r^2}\, . \\
  u^{(\beta)}(r) &\to& \frac{a_{02}}{a_0} r\, ,\quad w^{(\beta)}(r) \to \left( a_2 - \frac{a_{02}^2}{a_0} \right) \frac{3}{r^2}
  - \frac{r^3}{15} \, .
\end{eqnarray}
where $a_0$, $a_{02}$ and $a_2$ are the S-, S-to-D- and D-wave scattering
lengths (or ``volumes'', their dimensions are
  ${[{\rm length}]}^{l+l'}$ with $l$, $l'$ the initial
  and final orbital angular momenta, check Ref.~\cite{PavonValderrama:2005uj}
  for further details).
Here we will be interested in $a_0$, the calculation of which only requires
the isolation of the $\alpha$-type zero-energy wave function.
For simplicity, as $a_0$ is the only scattering length we will be
interested in, from now on we will refer to it as ``$a_t$'', i.e.\ the
triplet scattering length.

\subsection{Regulators and Regulator Dependence}
\label{sec:regulators}

Now we will compare the cutoff dependence with a series of regulators.
Within EFT the natural expectation is that if the auxiliary scale
(i.e.\ the cutoff) is harder than the light physical scales
in the systems, then observable quantities will be
independent of the regulator choice.
We advance that this will indeed be the case in the calculations
that follow.
Yet, it has also been argued that the residual cutoff dependence --- the cutoff
dependence before reaching the $R_c \to 0$ / $\Lambda \to \infty$ limit ---
provides information about the expected size of the subleading order
corrections.
That is, by carefully analyzing this residual cutoff dependence, we will be
able to determine the power counting of the EFT we are using~\cite{Long:2011qx,Long:2011xw,Long:2012ve,Griesshammer:2015osb,Valderrama:2016koj,Griesshammer:2020fwr}.
This second conclusion seems to be challenged by the calculations
we will show below.

We will compare five regularization schemes, representing a mixture of
the different local and nonlocal regulators that are frequently used
in the literature to overcome the divergence associated
with the short-distance behavior of the OPE.
In particular we will consider the following: (i) a boundary condition
at the cutoff radius $R_c$, (ii) a delta-shell regularization
for the contact-range potential  (iii) a local potential
regulator in r-space (a Gaussian), (iv) a semilocal regulator:
and (v) a non-local regulator, where the contact- and finite-range
potentials are regularized in p-space.
In the following lines we will explain each regulator in detail.

Independently of the regulator, we will consistently use the same
renormalization condition: we will calibrate the contact-range coupling
of the EFT to reproduce a deuteron binding energy of $2.2\,{\rm MeV}$.
Later, for checking the regulator dependence, we will calculate
the neutron-proton scattering length in the $^3S_1$ partial wave:
all regularization schemes will lead to the same scattering length
once the cutoff is hard enough ($a_t = 5.317\,{\rm fm}$), but
they will differ on how convergence to this value is achieved.

\paragraph{Boundary condition: }

A boundary condition exploits the fact that the Schr\"odinger equation is a
second order differential equation and its solution is thus determined by
two boundary conditions: the value of the wave-function and its derivative
at a given radius $r = R_c$.
Once we take into account that the normalization of the wave-function
can be determined \textit{a posteriori}, these two conditions are reduced to
a single boundary condition in the ratio between the derivative of
the wave-function and the wave-function itself.
To illustrate this let us consider the S-wave reduced Schr\"odinger equation
\begin{eqnarray}
  -u''(r) + 2 \mu V(r)\,u(r) = k^2\,u(r) \, ,
\end{eqnarray}
where $u$ is the reduced wave-function, $\mu$ the reduced mass,
$V$ the potential and $k$ the center-of-mass momentum.
The solution is completely determined
by the boundary condition
\begin{eqnarray}
  \frac{u'(r)}{u(r)} {\Big|}_{r = R_c} = D_k(R_c) \, ,
\end{eqnarray}
where $D_k(R_c)$ is the value of the log-derivative of
the wave function at $r = R_c$, which will be chosen
to fulfill a given renormalization condition
(e.g. reproducing a bound state or a scattering length).
The advantage of the boundary condition is that it is trivial to analyze
the residual cutoff dependence of observables,
which, if we choose $D_k(R_c) = D_0(R_c)$ (i.e. an energy-independent
  boundary condition), turns out to be~\cite{PavonValderrama:2007nu}
\begin{eqnarray}
  \frac{d \delta_k}{d R_c} \propto {| u_k(R_c) |}^2 \, .
\end{eqnarray}

While in a single-channel problem the choice of boundary conditions
is relatively simple (a condition on the log-derivative),
for a coupled-channel problem there are infinite choices,
a few of which can be consulted in Ref~\cite{PavonValderrama:2005gu}.
For simplicity we will choose the conditions
\begin{eqnarray}
  \frac{u'(R_c)}{u(R_c)} = D_{\gamma}(R_c) \quad \mbox{and} \quad w(R_c) = 0 \, ,
\end{eqnarray}
where $D_{\gamma}(R_c)$ is chosen as to reproduce the deuteron binding energy.
This boundary condition is the most direct coupled-channel extension of
the standard single-channel boundary condition.

\paragraph{Delta-shell: }

The next regulator is the delta-shell, in which the long-range potential
is cutoff for $r < R_c$
\begin{eqnarray}
  V^{\rm reg}_{\rm OPE}(r ; R_c) = V_{\rm OPE}(r)\,\rho(r; R_c)
  \quad \mbox{with} \quad
  \rho(r; R_c) = \theta(r - R_c) \, ,
\end{eqnarray}
and where the contact-range potential takes the form
\begin{eqnarray}
  \tilde{\chi}(r; R_c) = g(R_c) \, \delta_{l,0} ~ \delta_{l',0} \,
  \frac{\delta(r - R_c)}{4 \pi R_c^2} \, ,
\end{eqnarray}
which only acts on the $^3S_1$ partial wave ($l=l'=0$).
This contact-range potential is indeed equivalent to the following
energy-dependent boundary condition:
\begin{eqnarray}
  D_k(R_c) = \frac{m_N\,g(R_c)}{4 \pi R_c^2} + k \cot(k R_c) \, ,
\end{eqnarray}
where the energy dependence (a consequence of using a potential)
will generate results that are somewhat different to the energy
independent boundary conditions we have discussed above.
Despite this difference, many of the analytical results about residual cutoff
dependence with a boundary condition directly translate to
the delta-shell regulator.

\paragraph{Local regulator: }

For the local regulator, we will choose a Gaussian-like function.
Its specific form is
\begin{equation}
  \label{eq:local_regulator}
  \rho(r; R_c) = \left(1 - e^{-(r/R)^2}\right)^4~,
\end{equation}
where $R$ is the coordinate-space cutoff.
The local contact-like interaction used to fix the two-body binding energy is
\begin{equation}
  \label{eq:local_counterterm}
  \chi_{l,l'}(r) = g_n \, \delta_{l,l'} e^{-(r/R)^4} ~ .
\end{equation}
This counterterm is diagonal in angular momentum space, acting exclusively
in the ${}^3S_1-{}^3S_1$ and ${}^3D_1-{}^3D_1$ channels.
In the local case, there exists multiple ``branches'' of $g_n$, each
with a unique number of bound states~\cite{Beane_2001}.
Accordingly, $n$ denotes the total number of bound states.

\paragraph{Semilocal regulator: }

For the semilocal regulator the finite-range OPE potential is regulated locally
with the Gaussian-like function of Eq.~\eqref{eq:local_regulator}.
However, the contact-range interaction is regulated in p-space with a nonlocal
regulator:
\begin{equation}
  \label{eq:nonlocal_counterterm}
  \langle p' | \tilde{\chi} | p \rangle = g\, \delta_{l,0}~\delta_{l',0}~
  e^{-{(p' R_c / 2)}^4} \, e^{-{(p R_c / 2)}^4} \, .
\end{equation}
The advantage of this regularization over the local one is that now
there is only one branch for the coupling $g$, which greatly
simplifies the analysis of the residual cutoff
dependence.
Notice that our choice of nonlocal counterterm acts exclusively
in the ${}^3S_1$-${}^3S_1$ channel.

\paragraph{Nonlocal regulator: }

Finally for the nonlocal regulator, the finite-range OPE potential is regulated
in exactly the same way as the contact-range interaction:
\begin{eqnarray}
  \langle p' | V_{\rm OPE}^{\rm reg} | p \rangle =
  \langle p' | V_{\rm OPE} | p \rangle \,
  \tilde{\rho}(p' ; R_c)\,\tilde{\rho}(p; R_c) \, ,
\end{eqnarray}
where the regulator function is
\begin{equation}
  \label{eq:nonlocal_regulator}
  \tilde{\rho}(p; R_c) = e^{-{(p R_c/2)}^4} ~.
\end{equation}
Similarly, the contact takes the form given
in Eq.~(\ref{eq:nonlocal_counterterm}).
In the nonlocal case, where nonlocal regulators are applied to the OPE after the
Fourier transform, the cutoff input to the local regulator is $R/10$.
This removes the possibility of interference between the local and nonlocal
regulators.

\subsection{Results}
\label{sec:results}
\begin{center}
\begin{figure*}
  \includegraphics[width = 0.32\textwidth]{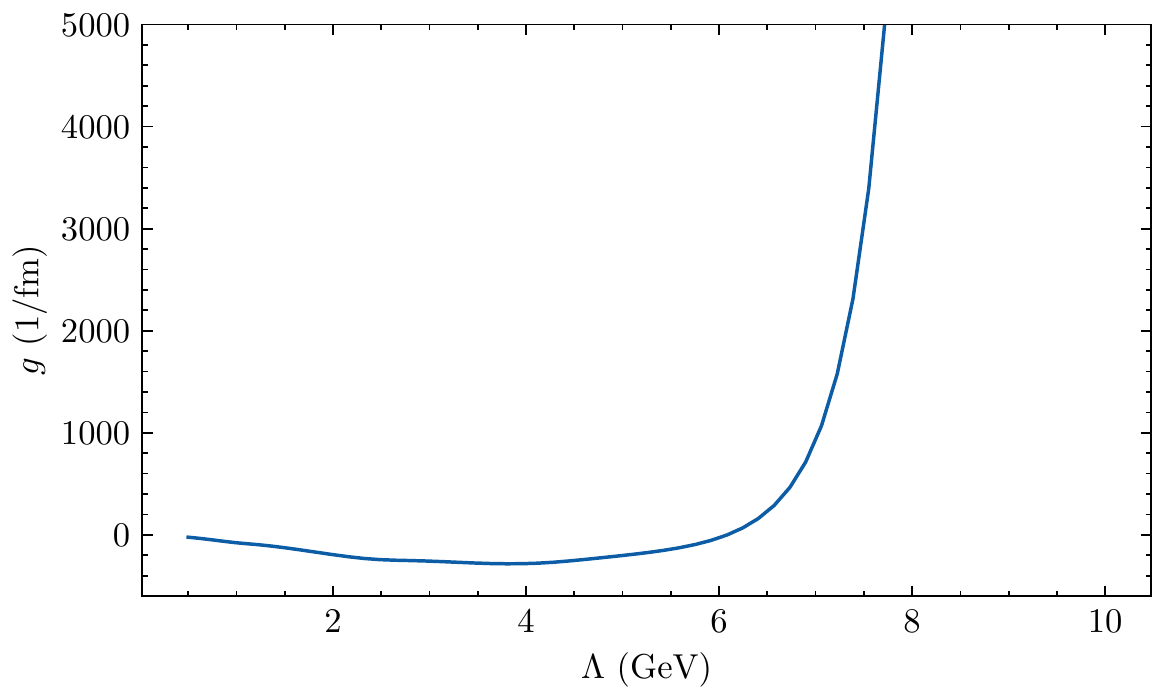}
  \includegraphics[width = 0.32\textwidth]{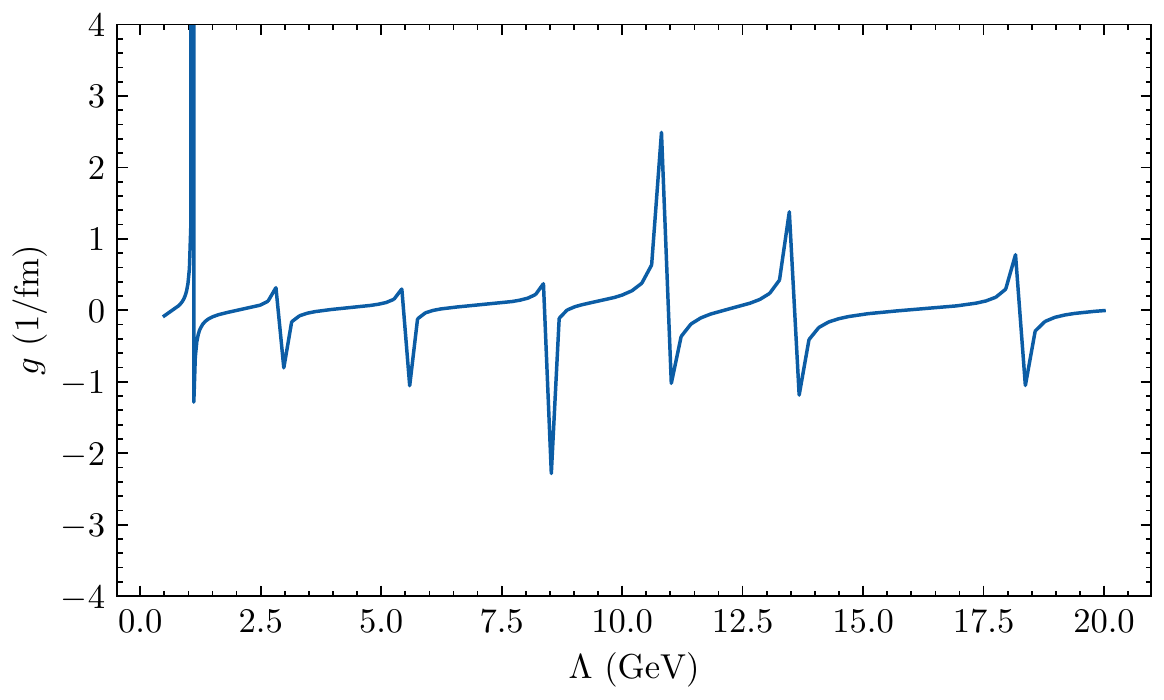}
    \includegraphics[width = 0.32\textwidth]{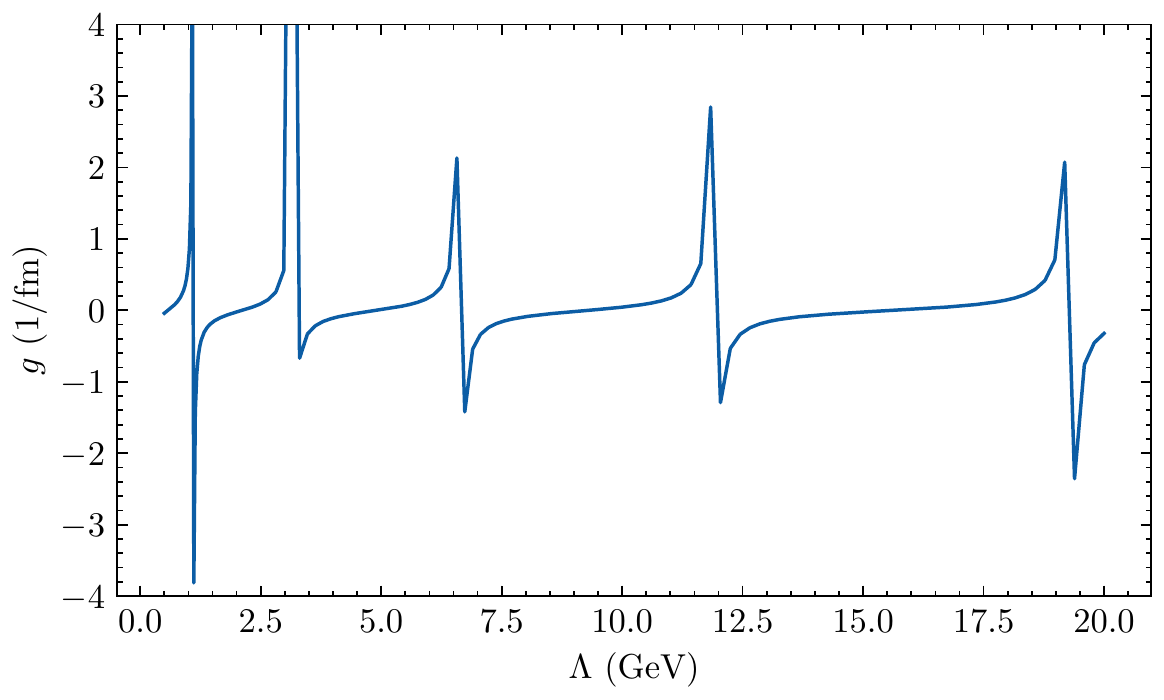}
  \caption{Left panel: The running of the counterterm coupling
    strength, $g_4$, for the local case. This branch corresponds to four
    bound states. Center panel: The running of the counterterm
    coupling strength, $g$, for the semi-local case. Each ``jump'' in
    $g$ as $\Lambda$ ($R$) increases (decreases) corresponds to the
    addition of a bound state. Right panel: The running of the
    counterterm coupling strength, $g$, for the nonlocal case. Each
    ``jump'' in $g$ as $\Lambda$ ($R$) increases (decreases)
    corresponds to the addition of a bound state.}
  \label{fig:rg_flow_local}
\end{figure*}
\end{center}

\begin{center}
  \begin{figure*}
    \centerline{
  \includegraphics[width = 0.47\textwidth]{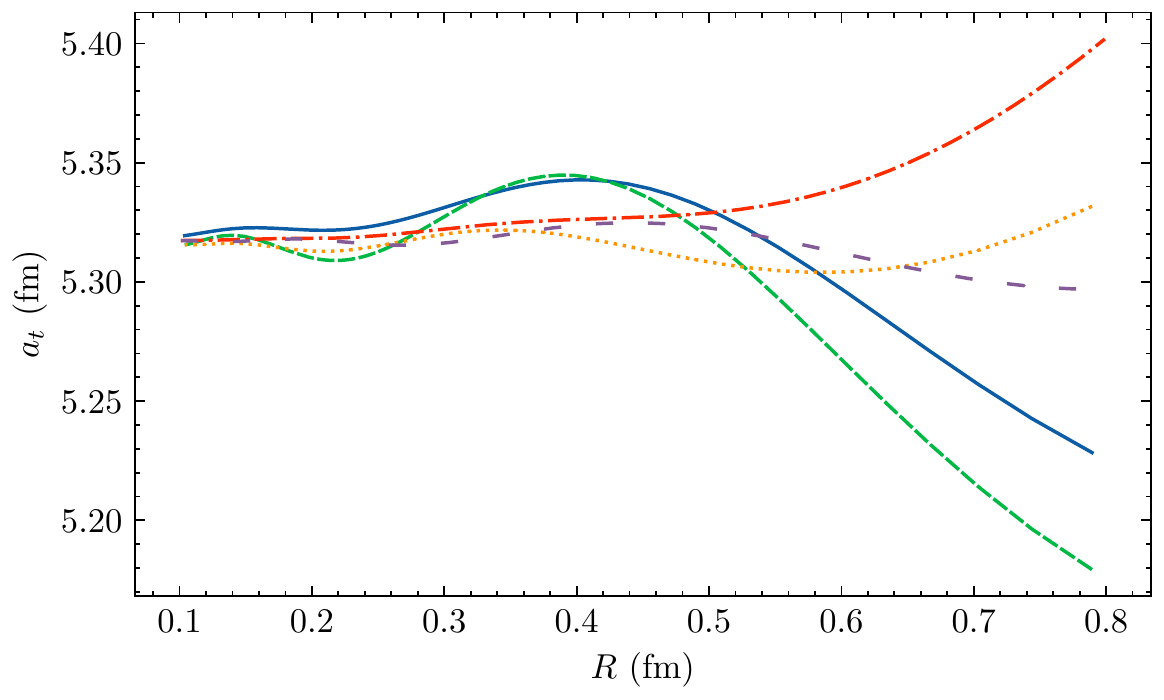}
   \includegraphics[width = 0.47\textwidth]{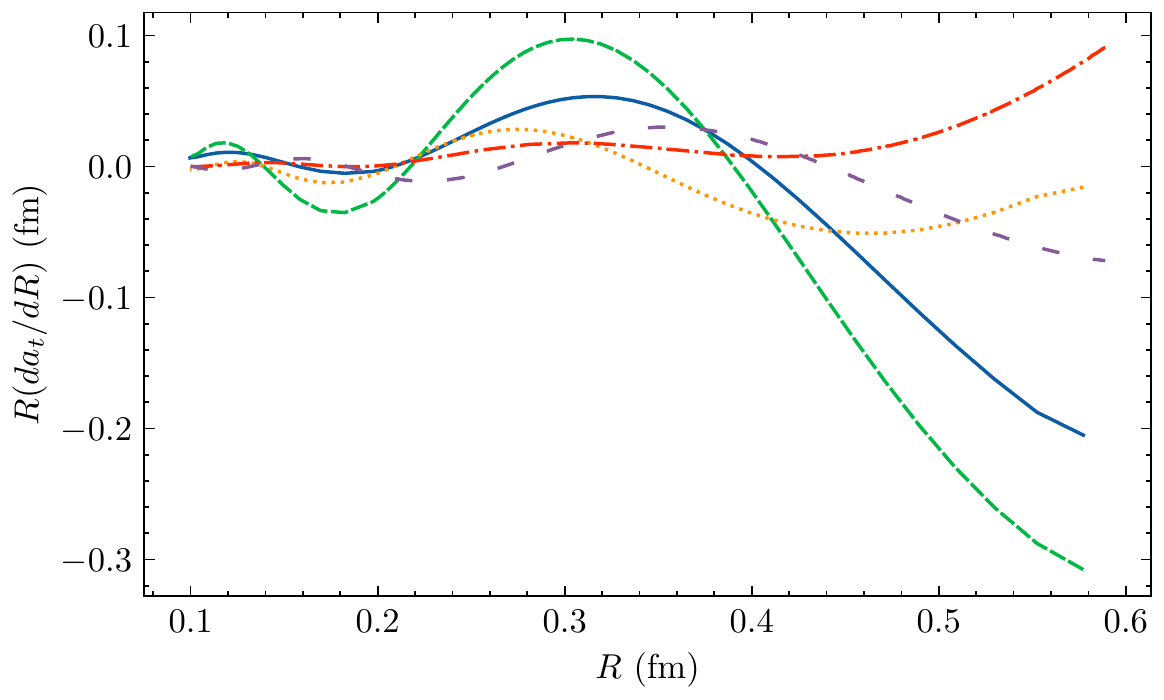}}
  \caption{Left panel: The triplet scattering length, $a_t$, is shown
    as the coordinate-space cutoff $R$ decreases. The solid, blue line
    is the nonlocally regulated calculation. The tightly dashed, green
    line corresponds to the semi-local case. The dotted, yellow line
    is the local calculation. The dash-dotted, red line is the
    boundary condition calculation. The loosely dashed, purple line is
    the $\delta$-shell calculation.  Right panel: The logarithmic
    derivative of $a_t$ with respect to the coordinate-space cutoff,
    $R$, is shown for the five different regulation/renormalization
    schemes.
  }
    \label{fig:scattering_length}
\end{figure*}
\end{center}

\paragraph{Renormalization Group Flow:}
The RG flows for the different regulation schemes are shown in the
panels of Fig.~\ref{fig:rg_flow_local}. The local RG flow, which
contains multiple branches, is shown for a branch with four bound
states, an arbitrary choice that permits fast and accurate
calculations.  It is important to note, as has been shown previously
in~\cite{Beane_2001}, that this is only one of infinitely many
``branches''.  Other branches can be followed, each corresponding to a
unique number of bound states.  A qualitative comparison of the local
RG flow to the semi-local or nonlocal RG flows already demonstrates
the regulator dependence of the limit cycle.
\paragraph{Spin-Triplet Scattering Length:}
The spin-triplet scattering length is shown in
Fig.~\ref{fig:scattering_length} for five different
regulation/renormalization schemes.
One important feature of the results shown is the agreement of the
asymptotic values between the different schemes. The follow-up
question is immediately the means by which that asymptotic value is
reached.  One might expect that the precise form of $a_t(R)$ is
regulator-dependent.  What may not be so obvious is how consistent the
extraction of higher-order corrections is.

\paragraph{Higher-Order Corrections Analysis:}

To analyze the higher-order corrections, we take the logarithmic
derivative of $a_t$ with respect to the coordinate-space cutoff,
$R(da_t/dR)$.  The results for the five different regulation schemes
are shown in the right panel of Fig.~\ref{fig:scattering_length}.

The quantity $R(da_t/dR)$ emphasizes the oscillatory behaviors of
$a_t$'s convergence in each scheme.  In particular, one can readily
observe differences in that oscillatory behavior --- in amplitude and
period --- across the different schemes.  In fact, the only
consistencies are the oscillatory behavior and the asymptotic approach
to zero.

As has been done in the past~\cite{Song:2017}, one could assume that the
convergence of a given observable follows a typical expansion of the form
\begin{equation}
  \label{eq:observable_expansion}
  \mathcal{O}(R) = \mathcal{O}_\infty + \sum_{i=1}^\infty c_i (qR)^i~,
\end{equation}
where the $\mathcal{O}_\infty$ is the asymptotic value, $c_i$ are the
coefficients of the expansion, and $q$ is the relevant momentum scale.
Each regulation/renormalization scheme is a different manifestation of
the same LO EFT.  As such, one might expect that a truncation of
\eqref{eq:observable_expansion} after the first term ought to provide
a reasonable description of $a_t(R)$ for each of the different cases.
But as is clearly shown in Fig.~\ref{fig:scattering_length}, such
a simple analytic form is not sufficient.

To overcome this added complexity, we considered a modified LO expansion
\begin{equation}
  \label{eq:modified_expansion}
  \mathcal{O}(R) \approx \mathcal{O}_\infty + c~f(R) R^n~,
\end{equation}
where $f(R)$ is some oscillatory function that is periodic
in $R$ or some power of $R$. Taking the general form
\begin{equation}
  \label{eq:f_R}
  f(R) = C_1\cos{(\omega R^m + \phi)} + C_2~,
\end{equation}
we constructed an expression for the logarithmic derivative and fit the
unknown parameters as in~\cite{Odell:2019wjq}.
It is worth noticing that the previous form of the residual cutoff
dependence is analogous to the one calculated for boundary condition
regularization in Ref.~\cite{PavonValderrama:2007nu}.
However, fitting to this form has proven to be unreliable and
very sensitive to the range of $R$ values over which the fit is performed.

\begin{center}
  \begin{figure*}
    \centerline{
  \includegraphics[width=0.46\textwidth]{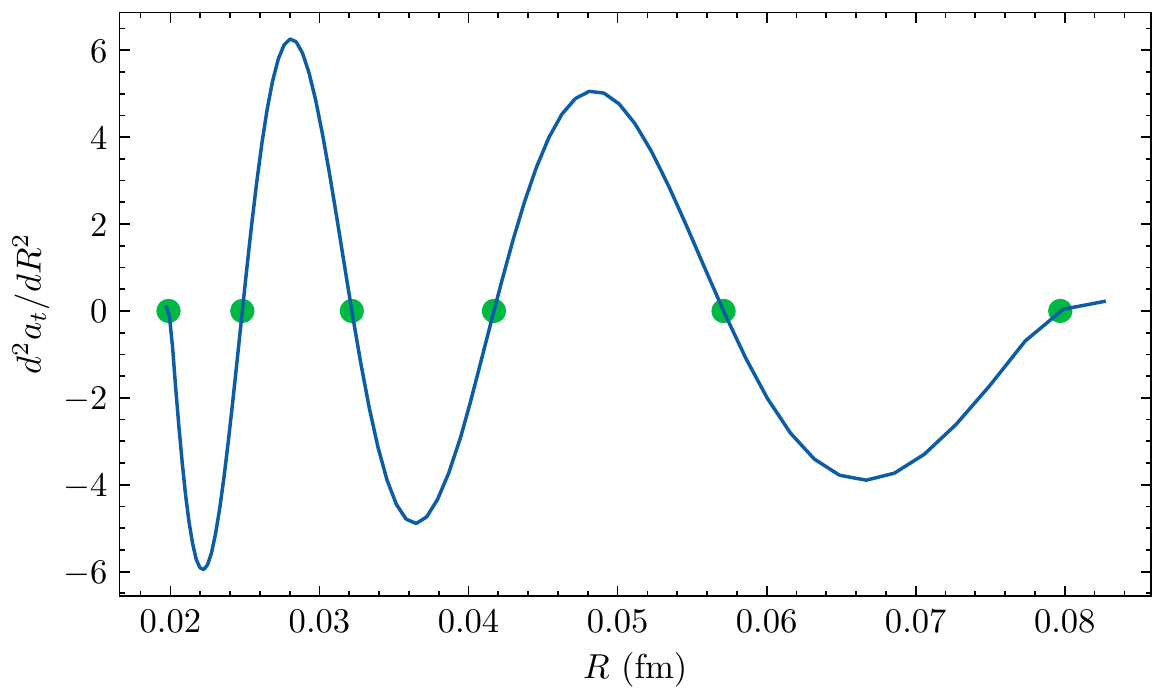}
  \includegraphics[width=0.46\textwidth]{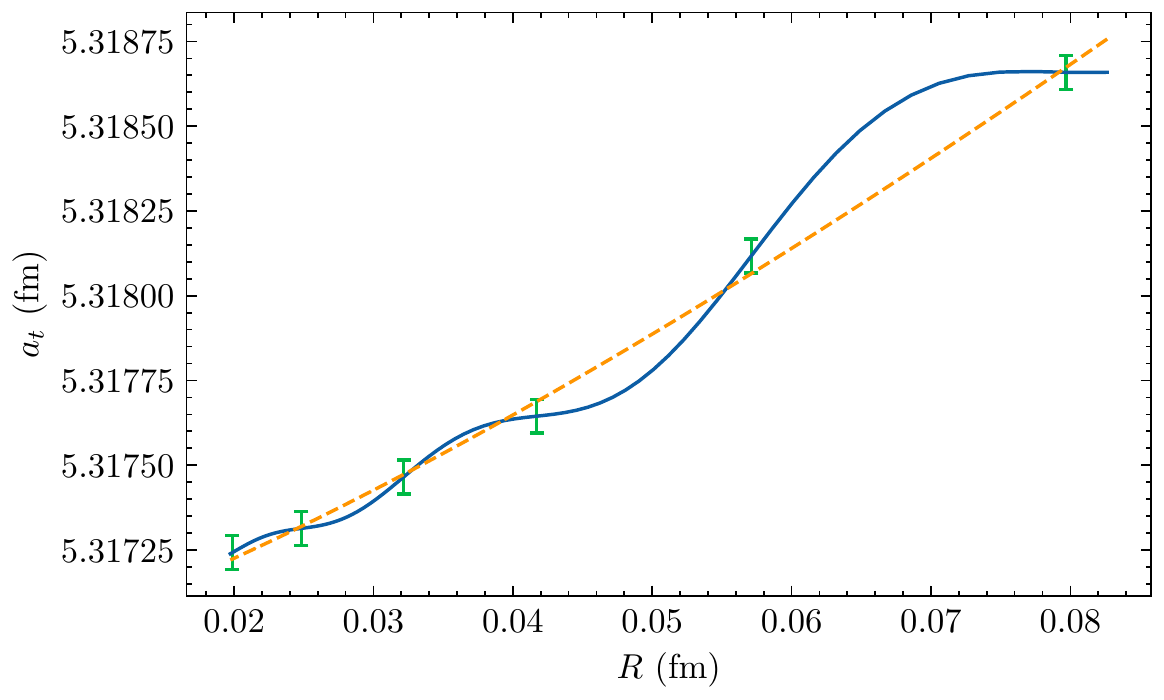}
  }
  \caption{Left panel: $d^2a_t/dR^2$ is shown as a solid, blue
    line. Green circles highlight the roots of the function. Results
    are generated with the nonlocal regulation scheme.  Right panel:
    $a_t(R)$ is shown as a solid, blue line. The locations of the
    roots of $d^2a_t/dR^2$ and the associated error bars are indicated
    with green error bars. The resulting fit to Eq.~\eqref{eq:h_R} is
    shown as a dashed, orange line.}
  \label{fig:d2a_roots}
\end{figure*}
\end{center}

Instead, we found it much more effective to consider points in the $R$
dependence of $a_t$ where $f(R)$ repeats itself.
To do this, we study the roots of $d^2a_t/dR^2$.
In some cases, a constant was subtracted from $d^2a_t/dR^2$ to extract more
intersections.
As an example, the results for the second derivative of $a_t$ found using the
nonlocal scheme are shown in Fig.~\ref{fig:d2a_roots}.
Once generous error bars attributable to numerical stability are assigned, it is
a straightforward process to get reliable estimates for $n$ by fitting $a_t(R)$
evaluated at the roots $R_i$ (with $i = 1,2, \dots$) of $d^2a_t/dR^2$ to
\begin{equation}
  \label{eq:h_R}
  h(R_i) = \mathcal{O}_\infty + c R_i^{n_{\rm{opt}}}~,
\end{equation}
where $\mathcal{O}_\infty$, $c$, and $n_{\rm{opt}}$ are treated as fit parameters.
The results for the nonlocal case are shown in the right panel of
Fig.~\ref{fig:d2a_roots}.
For this particular case, there are stable results over a wide range of
small $R$ values.
This allows us to include six values of $a_t(R)$ in the fit to
Eq.~\eqref{eq:h_R}.
The results for all five regulation/renormalization schemes are summarized in
Table~\ref{tab:n_opt}.
In some cases, numerical stability limited the number of data points to fit
(for instance, in the boundary condition and delta-shell cases),
and the uncertainty estimates for $n_{\rm{opt}}$ reflect that limitation.
Specifically, for the semi-local and local cases, in addition to the covariance
returned by the least-squares fit to Eq.~\ref{eq:h_R}, uncertainty estimates for
$n_{\rm{opt}}$ take into account the dependence of $\chi^2/\nu$.
To establish that dependence, we fixed $n_{\rm{opt}}$ over a range of values and fit only
$\mathcal{O}_\infty$ and $c$.
The estimates from the covariance matrix of the least-squares fit agree well
with values of $n_{\rm{opt}}$ that produce $\chi^2/\nu$ values equidistant from one.

It appears from the variety of results, extracting the order of higher-order
corrections for cutoff dependence is not a reliable practice.
However, one may still be able to make reasonable comparisons between two- and
three-body results obtained with the same regulators as was done in
\cite{Odell:2019wjq}.
\begin{table}
  \caption{Upon fitting $a_t(R)$ to Eq.~\eqref{eq:h_R}, the optimal value of
    $n_{\rm{opt}}$ and its associated uncertainties are estimated primarily from
    the covariance matrix of the least-squares fit. Those values are summarized
    here for all five regulation/renormalization schemes. The range of $R$
    values over which the fit was conducted is also given for each scheme.}
\label{tab:n_opt}       
\begin{tabular}{llllll}
\hline\noalign{\smallskip}
 ~ & Nonlocal & Semi-local & Local & BC & $\delta$-Shell  \\
\noalign{\smallskip}\hline\noalign{\smallskip}
 $n_{\rm{opt}}$ & 1.3~$\pm$~0.2 & 3.4~$\pm$~1 &  6$\pm$3 &  1.4$\pm$0.35 & 2$\pm$0.9   \\
 $R$ range (fm) & [0.02, 0.084] & [0.03, 0.06] &  [0.07, 0.2] &  [0.17, 0.45] &
 [0.15, 0.5] \\
\noalign{\smallskip}\hline
\end{tabular}
\end{table}

Finally, it is worth noticing that for a single-channel problem the
boundary condition method allows one to calculate the cutoff dependence
semi-analytically~\cite{PavonValderrama:2007nu}
\begin{eqnarray}
  \lim_{R \to 0}\,\frac{d O}{d R} \Big|_{\rm BC} \propto {\left| u(R) \right|}^2
  \propto R^{3/2}\,f(R) \, ,
\end{eqnarray}
with ``BC'' indicating ``boundary condition'' and where we have
particularized for a potential behaving as $1/r^3$ at short-distances.
If this result is translatable to the coupled-channel deuteron case,
it would indicate that at small enough $R$, the boundary condition
method should yield $n_{\rm opt} = 2.5$, which is not compatible with
Table~\ref{tab:n_opt}, implying that the cutoff range $R$ used is not
small enough as to observe the expected power-law behavior.
Numerical limitations make it impractical to study the boundary
condition at shorter $R$, which leads us to the following observation:
even though in this case we know that the power-law behavior of the
residual cutoff dependence will eventually approach a certain value,
practical EFT calculations might still be unable to reach the cutoff
range for which this happens.
This indeed represents another important constraint to the usefulness
of residual cutoff dependence.

\begin{center}
\begin{figure}
  \includegraphics{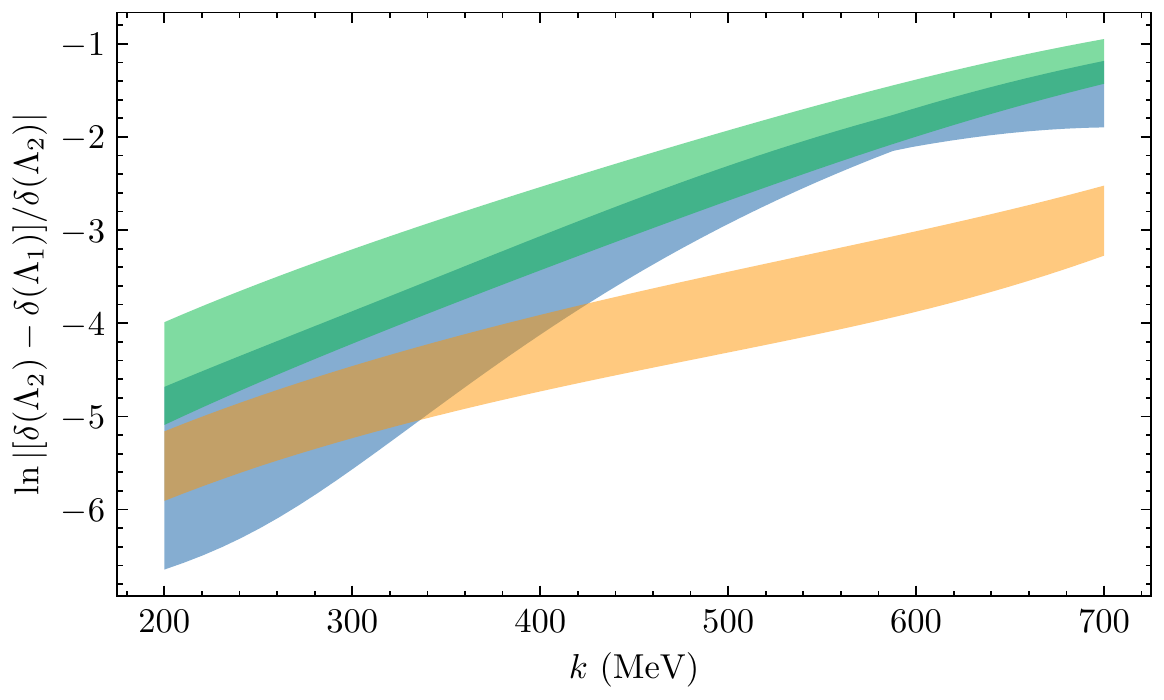}
  \caption{The logarithm of the absolute value of the relative difference
    between the phase shifts calculated at $\Lambda_1$ and $\Lambda_2$ is shown
    as a function of the scattering momentum. $\Lambda_1$ is fixed at 714 MeV.
    $\Lambda_2$ varies up to 3.3 GeV. The nonlocal results are shown in blue,
  semi-local in green, and local in orange (color online).}
  \label{fig:griesshammer_plot}
\end{figure}
\end{center}

\paragraph{Energy dependence analysis of the phase shifts:}
An alternative approach to the analysis was proposed by Griesshammer
in Ref.~\cite{Griesshammer:2015osb} who suggested to consider the
ratio of the scattering phase shift calculated at two different
regulator values as a function of the momentum scale. The slope of the
resulting momentum dependent curve is then supposed to inform on the
correctness of the underlying powercounting employed in an effective
field theory.
In Fig.~\ref{fig:griesshammer_plot} we show the results of this
analysis obtained by calculating the phase shifts at two different cutoffs,
$\Lambda_1$ and $\Lambda_2$ (corresponding to two different values of $R$).
$\Lambda_1$ is fixed at 714 MeV.
$\Lambda_2$ varies up to 3.3 GeV.
The band is generated by overlaying all of the curves for each value of
$\Lambda_2$ and taking the maximum and minimum values at each value of $k$.

The most important aspect of Fig.~\ref{fig:griesshammer_plot} is the apparent
discrepancy between the slopes of the different bands.
Even within the same regulation scheme, the slope is not constant.
Secondarily, the range of $k$ was chosen to extend beyond 600 MeV to study the
breakdown scales.
As discussed in \cite{Griesshammer:2015osb}, notable changes in the slope ought
to correspond to the breakdown scale of the theory, previously estimated between
450 and 600 MeV.
While there are changes, a consistent inference of that breakdown scale can not
be made from the bands shown in Fig.~\ref{fig:griesshammer_plot}.

\section{Summary}
\label{sec:summary}
In this work, we showed that regulator dependence of observables
calculated with the leading order chiral potential is significantly
more complex than advertised in the existing literature. We find that
the residual regulator dependence of an observable is generally
$f(\Lambda) \left(Q/\Lambda\right)^n$ where $f(\Lambda)$ is an
oscillating function whose period and amplitude depends on the
specific form of the regulator. We emphasize that an extraction of the
power of the residual cutoff dependence is seriously complicated by
this feature and thereby also the extraction of the power $n_{\rm{opt}}$ of the
residual cutoff dependence. Furthermore, our analysis seems to
indicate that also the power $n_{\rm{opt}}$ of the cutoff dependence itself
depends on the regulator employed.

The standard expectation for the residual cutoff dependence is $n_{\rm opt} =
2.5$ (in the boundary condition method),
which leads to the conclusion that subleading corrections
in the deuteron enter at ${\rm N^{5/2} LO}$ (where this fractional
power counting is often simplified to ${\rm N^2 LO}$ or ${\rm N^3LO}$
depending on authors).
While the $\delta$-shell and perhaps the semi-local and local
regulators lead to a residual cutoff dependence compatible with this
expectation, similar conclusions can not be drawn for the nonlocal
and, surprisingly, boundary condition schemes.
The previously mentioned $n_{\rm opt} = 2.5$ expectation can be
interpreted as being compatible with the original Weinberg counting,
at least for the particular case of the deuteron channel where
subleading corrections should enter at ${\rm N^2LO}$.
In contrast, nonlocal and boundary condition schemes point towards a
smaller power-law dependence of $n_{\rm opt} \approx 1.5$, which will
indicate that subleading order corrections enter much earlier than
originally expected, while the local regulator could be interpreted as
indicating the contrary, that is, subleading orders being pushed to
higher orders.
This is surprising at first sight as one naively expects that the structure of
the EFT expansion determines the power law behavior of the residual corrections.
We suspect that interplay between the local components of the OPE potential and
the regulator are responsible for our observations.
These findings call therefore for the need of a better understanding of
regulator effects in EFTs as it needs to be assessed whether residual cutoff
dependence can still be considered a tool in the analysis of the power counting
of EFT\@.
Our results show furthermore that the estimation of uncertainties by
variation of the regulator scale cannot be considered reliable unless
$f(\Lambda)$ is known \textit{a priori}.
Our results demonstrate that variations around larger
values of the coordinate space cutoff, $R$, can lead to dramatically
different estimates of EFT uncertainties.

However, the analysis presented here is still subject to a series of known
unknowns: improved numerical accuracy (particularly in the determination of the
power-law exponent), dependence on the choice of branches for local regulators,
etc., which will deserve further attention in the future.
It remains possible that higher-order correction analyses conducted at
sufficiently large $\Lambda$ produce consistent results across different
regulation/renormalization schemes.
At present, in light of the current results, even if this optimistic
view is materialized, its value will mostly be academical: practical
EFT calculations are unlikely to be able to explore the cutoff
windows, branch-dependence and other related factors that are
necessary to uncover the underlying cutoff dependence.
This puts strong limitations on how residual cutoff dependence can
actually be of use within EFT, except for certain easy examples.

\begin{acknowledgements}
  We thank Ubirajara van Kolck for useful discussions. This work has
  been supported by the National Science Foundation under Grant
  No. PHY-1555030, the Office of Nuclear Physics, U.S. Department of
  Energy under Contract No. DE-AC05-00OR22725, and the National
  Nuclear Security Administration No.~DE-NA0003883.
  M.P.V. has been supported by National Natural Science Foundation of China
  under Grants No. 11735003 and 11975041, the Fundamental
  Research Funds for the Central Universities and the Thousand
  Talents Plan for Young Professionals.
  M.P.V. would also like to thank the IJCLab of Orsay, where part of this work
  has been done, for its long-term hospitality.
\end{acknowledgements}

\bibliographystyle{apsrev}       

%
%

\end{document}